# Thermal emission of hydrogenated amorphous silicon microspheres in the mid-infrared

*Roberto Fenollosa,\* Fernando Ramiro-Manzano*

Instituto Universitario de Tecnología Química, CSIC-UPV, Universitat Politècnica de València, Av. dels Tarongers, València 46022, Spain

E-mail: rfenollo@ter.upv.es

**Abstract**

Hydrogenated amorphous silicon microspheres feature a pronounced phononic peak around 2000 cm$^{-1}$ when they are thermally excited by means of a blue laser. This phononic signature corresponds to vibrational modes of silicon-hydrogen bonds and its emitted light can be coupled to Mie modes defined by the spherical cavity. The signal is apparently quite stable at moderate excitation intensities although there appeared some signs pointing to hydrides bonds reconfiguration and even hydrogen emission. Above a certain excitation threshold, a phase change from amorphous to poly-crystalline silicon occurs that preserves the good structural quality of the microspheres.

# 1. Introduction

Hydrogenated amorphous silicon (a-Si:H) is a material that has raised the interest of many researchers in the past years. It constitutes an alternative to crystalline silicon in the field of semiconductors and particularly for photovoltaic cells with several advantages such as the possibility of decreasing the cells thickness yet with keeping a high photonic absorption. However, it has several disadvantages such as the short carrier lifetime because of the high number of structural defects, although it is precisely the presence of hydrogen atoms that helps passivate them and improve the optoelectronic performance of the material [1]. In fact, this passivation effect of hydrogen has been successfully utilized for improving the efficiency of solar cells and it constitutes an important research area in this field [2,3].

a-Si:H can be synthesized by different methods. For example, hydrogen addition during sputtering depositions and glow discharge of silane [4,5]. In addition, the variation in the deposition parameters can lead to a variety of amorphous silicon materials with different amounts of dangling bonds and hydrogen atoms as well as silicon hydrides configurations: Si-H, Si-$H_2$ and Si-$H_3$. Therefore, there is, under the umbrella of a-Si:H, a variety of materials with different optical and electrical properties. All this makes a-Si:H be a versatile material that cannot only work as a base platform for developing optoelectronic devices such as solar cells and detectors, but it can also be combined with c-Si to form heterojunction solar cells, and additionally, it has been considered a promising platform for nonlinear photonics [6].

An important feature of a-Si:H is its instability against light radiation mainly in the visible range, particularly the decrease of the photoconductivity and the dark conductivity by light soaking. This is called the Saebler-Wronski effect [7]. It is in fact a reversible photoelectronic effect because it was shown that annealing of the material above 150 ºC could reverse the process. This phenomenon has not been fully understood so far, and in fact some researchers have recently reported the opposite effect, i. e. an increase of conductivity, occurred under some circumstances with light irradiation [8]. In any case this property is not necessarily a disadvantage, rather it could be applied for developing memory type devices like for instance neuromorphic building blocks which require reversible states.

Here, we report a study of amorphous silicon in the form of microspheres. Because of their fabrication process, which is based on di-silane as a precursor gas, these microspheres contain a large amount of hydrogen atoms bounded to the silicon matrix. Therefore, they are more properly called hydrogenated amorphous silicon microspheres. We reported several years ago about their synthesis [9]. Other noteworthy approaches producing monodisperse microspheres have been reported as well [10].

Concerning the characterization of a-Si:H, optical absorption in the mid infrared has been one the most utilized techniques. In contrast, it should be stressed that here we study this material from a different point of view, namely by measuring and analysing the thermal emission of microspheres, one by one. Rather than attempting to be exhaustive in explaining the complex behaviour of this material, we consider this work as a first step in exploring the possibilities as base material for future applications. In this regard, our approach has important points to be mentioned: Firstly, we have studied the thermal emission by increasing the temperature of the microsphere under study, and this is accomplished by laser irradiation. In this first stage, the temperature increase of the amorphous microsphere is assumed to be proportional to the light intensity exposure. As we reported in previous studies [11] such increase of temperature occurs relatively fast, on the order of ten microseconds because of the micrometer size of the particles. Secondly, this light-induced thermal process ends up in a phase transition allowing us to to achieve a crystalline phase, in the same way as it was used many years ago for crystallizing a-Si:H films [12,13]. Thirdly, in addition to the midIR characterization, complementary light-scattering measurements in the VIS and nearIR have been performed at room temperature. This allows testing if there have been structural changes during the thermal experiments and before the phase change because the measurements yield peaks corresponding to Mie resonances that strongly depend on structural parameters such as sphere diameter and refractive index.

## 2. Experimental

The synthesis of a-Si:H microspheres has been detailed elsewhere. Basically, it consists of introducing di-silane in a close reactor at a low pressure, typically about 10 kPa, and heating it at temperatures between 400 ºC and 500 ºC. This way, particles having a highly spherical shape nucleate and grow in the gas phase by means of a complex process [14], and finally fall down onto a substrate previously introduced in the reactor. It should be pointed out that because the synthesis temperature is lower than 600 ºC which is considered as the temperature where silicon crystallizes [15,16], the obtained particles have an amorphous nature. After the synthesis, individual microspheres with diameter around 3-4 µm were transferred to a glass substrate by a pick and place procedure for performing optical measurements.

Thermal emission measurements were performed on single microspheres by using a setup described in [17]. The sample holder is placed in a piezo stage attached to a mechanical positioner for long range and fine and stable positioning. A single particle is heated by means of a blue laser ($\lambda$=405 nm) focalized by an optical objective. Typical power intensities are between 5 and 15 mW. This pump signal is chopped at certain frequency (typically around 600 Hz). The emitted signal, after being collected by a Cassegrain objective, passes through a Michelson interferometer which is formed by a beam splitter made of ZnSe and two retroreflectors. One of the optical paths, corresponding to one of the interferometer arms, is varied by means of a linear stage. The mechanical actuator consists of a stepper motor attached to a mechanical gearbox (1:100). The resulted signal is acquired by a liquid nitrogen cooled MCT and subtracted by a lock-in amplifier. A secondary chopped (HeNe) red laser passes thought the interferometer and is collected by a Si-detector, attached to another lock-in amplifier. This allows for a precise determination of the variation of the optical path.

Optical scattering measurements in the visible range were performed at 90º of direction using a similar setup than that reported in [18]. It consists of illuminating with white light, in the same way as the blue light for thermal emission. However, instead of Cassegrain collection, a nearIR lens-based objective is employed. The signal is finally acquired by a monochromator attached to liquid-nitrogen-cooled CCD array detector.

## 3. Results and discussion

a-Si:H microspheres have in principle an external appearance which is similar to that of poly-crystalline ones, in particular with regard to size distribution, roughness and spherical perfection (see figure 1). However, their internal structure as well as their optical properties in the visible and near infrared ranges are different [18]. In what follows, we are going to see that differences are even more pronounced in the mid infrared range, specially when they are submitted to a thermal excitation process.

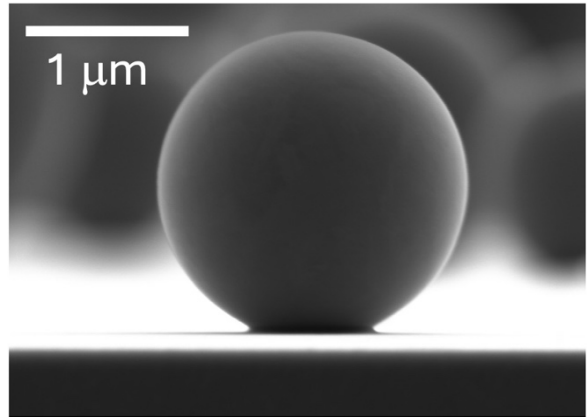

Fig1: Field Emission Scanning Electron Microscope (FESEM) image of a typical silicon microsphere utilized for thermal emission experiments.

*3.1 Thermal emission spectra of single a-Si:H silicon microspheres.*

Recently we have studied the thermal emission properties of silicon microspheres in the near [11] and in the mid infrared [17] ranges. In those studies, the type of silicon was poly-crystalline and the spectra showed several pronounced peaks that are originated from the coupling of free carrier (*FC*) emission to Mie resonant modes. In order to observe such resonances a sufficient pump intensity was required. Bellow this excitation, the structured thermal emission was under our limit of detection.

In principle, when an a-Si:H microsphere is irradiated by a blue laser, its temperature is expected to increase because of the large absorption at that wavelength and its relative thermal isolation from the surrounding environment, similarly to what occurs with a poly-crystalline one. However, at a certain excitation intensity, below the threshold where *FC* emission coupling to Mie resonances is visible, we observed several signatures which are typical of bond vibrations of Si hydrides [19] (see Fig. 2 (a)). They strongly resemble the absorption bands of amorphous silicon prepared by reactive sputtering in gas mixtures of $H_2$. The main peak is in fact a doublet with wavenumbers around 2000 cm$^{-1}$ and 2070 cm$^{-1}$, being the weaker shoulder the one at lower energy. This doublet is revealed by the asymmetry and broadening of the peak, and in principle it could indicate the presence of different types of hydrides, associated with vibrations in the amorphous matrix, specifically stretching of the Si-H bonds ($w_1^S$) for the lowest energy band and stretching of the Si-H$_2$ bonds (w$_2^S$) for the highest energy band. However, signatures corresponding to Si-H$_2$ bending modes (w$_2^B$ and w$_3^B$) should appear as well around 845-890 cm$^{-1}$ because this vibration mode requires the contribution of at least two hydrogen atoms. However, in our experiments they show a very low intensity. In fact, this spectrum has not been corrected for the system sensitivity precisely for being able to visualize them, because such a correction renders an almost negligible signal when plotted alongside the short wavenumber peaks. Therefore, we hypothesize that the shoulder at higher energy is rather produced by vibrations of Si-H bonds at the surface of micro or nano-voids existing in the amorphous matrix [20].

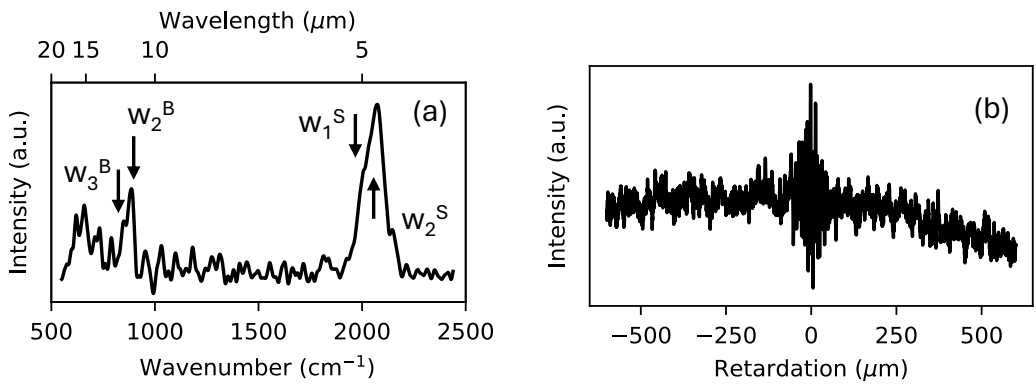

Fig. 2: (a) Thermal emission spectrum of an amorphous silicon microsphere. (b) Corresponding interferogram.

Figure 2 (b) shows the measured interferogram from which the spectrum of fig 2(a) was calculated by Fourier Transform. The maximum retardation is ± 600 µm thus providing a resolution of 17 cm$^{-1}$ and the whole measurement took 77 minutes. The interferogram illustrates the stability of the hydrides phonon emission during such a long period. The progressive decrease of the signal after the zero retardation point is caused by misalignments of the optical system because the signal could be recovered after proper adjustments of the setup.

In contrast to the apparent stability of the interferogram, the optical scattering in the visible-near infrared range indicated that important structural changes had taken place during these thermal emission experiments. In particular, the scattering measurements yield pronounced peaks that are associated with Mie resonances. Therefore, they are very sensitive to changes in the sphere diameter and refractive index. Figure 3 shows the optical scattering spectra at room temperature for an amorphous silicon microsphere before any irradiation (bottom curve) and after two successive irradiations similar to the one used to take the interferogram of fig 2 (b).

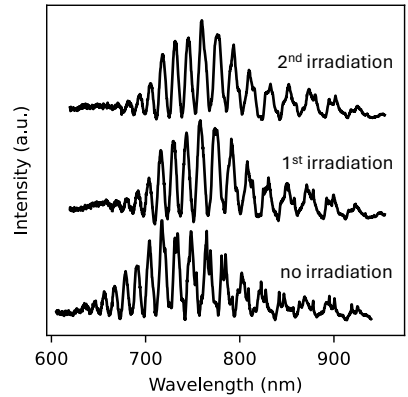

Fig 3: Optical scattering of an amorphous silicon microsphere performed before any irradiation and after two successive irradiations of 77 minutes for measuring the phonon emission of silicon hydrides.

The biggest differences with regards to peaks position and shape occurred between the spectrum corresponding to the non-irradiated microsphere and that corresponding to the 1$^{st}$ irradiation (the 2$^{nd}$ irradiation produced minimum differences with respect to the previous one). Although we have quantified changes in the sphere diameter and refractive index during the transition from amorphous to polycrystalline phases in silicon microspheres during an annealing process [18], the spectra of fig. 3 indicates that such changes start to occur at much earlier stages, before the phase transition. This changes in the optical scattering spectra may be attributed to migration of hydrogen atoms in the amorphous Si matrix and perhaps rearrangements of this matrix as well [21,22]. However, it seems that microspheres reach a kind of stable condition after the first irradiation process, while still keeping their hydrogenated amorphous nature.

*3.2. Conversion of a-Si:H into poly-crystalline silicon microspheres by laser irradiation.*
The increment of temperature in a-Si:H microspheres manifests firstly as MidIR phonons emission of Si hydrides as we have seen above. Because the temperature of a microsphere is supposed to increase with the intensity of the excitation laser, there should be a threshold where a phase transition from amorphous to poly-crystalline follows. This is an irreversible change that occurs around 600 ºC according to the literature [15,16]. Here, we have studied this process using the thermal emission of the microspheres to witness the phase transition, that is expected to manifest as a radical change in the spectral features.

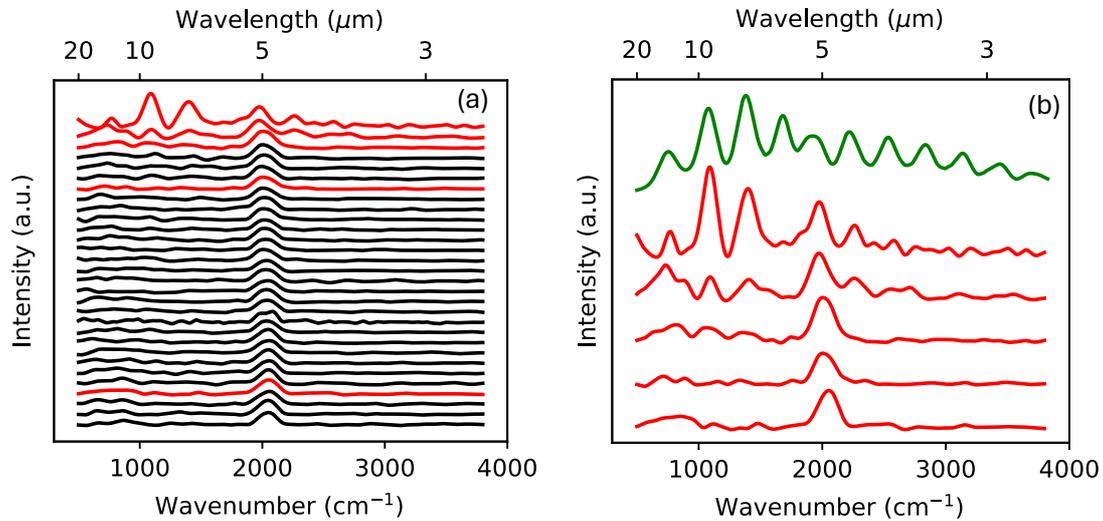

Fig 4: (a) Emission spectra of an amorphous silicon microsphere at increasing laser excitation intensities from bottom-up. Red spectra have been plotted in (b) in the same order for better visualizing the phase change from amorphous to poly-crystalline silicon. The green curve corresponds to a Mie theory based simulated emission spectrum of a crystalline silicon microsphere of 3600 nm in diameter at 600 ºC. Appropriate convolution has been realized in order to account for the finite resolution of the measurement.

Figure 4(a) shows the measured spectra for an a-Si:H microsphere recorded at increasing excitation steps (from bottom-up), between 0.2 and 0.6 mW, from 5 to 15 mW approximately. They were measured at low resolution (100 cm$^{-1}$), i.e. at short retardation lengths (± 100 μm), for increasing the speed in recording the process. At first glance, no appreciable changes occur in the initial stages, from the low to, say, medium-high power spectra (before any phase transition). The plots feature a prominent peak, around 2000 cm$^{-1}$, representing the main visible feature. However, a detailed data analysis of these spectra, shown in figure 5, reveals that (i) the the emission peak center (red dots) shifts progressively to shorter wavenumbers with the increase of the pump power; (ii) the peak intensity (black dots) remains quite stable, except for the case of the highest excitation intensity, in spite of the fact that the temperature of the microsphere is supposed to increase as the excitation intensity increases. Because such shift covers the spectral positions of the individual peaks of the doublet (around 2000 cm$^{-1}$ and 2070 cm$^{-1}$), we think that it could be produced by the rearrangement or thermal release of hydrogen atoms

[22] probably from the voids. This is, however, a qualitive interpretation and more research in this regard should be performed in order to better asses this hypothesis.

Figure 4 (b) includes a selected number of spectra from fig. 4(a) (red curves) and it illustrates the change of scenario at a certain power, characterized by the emergence of new peaks that we associate with Mie resonances. This was corroborated by a simulated spectrum based on Mie theory (green curve) that corresponds to a crystalline silicon microsphere with 3600 nm in diameter at 600 ºC. We applied appropriate convolution for accounting for the finite resolution of the measurement that broadens the peaks (see fig. 6(a) and (b) for similar simulations at higher resolution). The absence of peaks at shorter wavelengths in the experiments, in comparison with theory, is because of the low sensitivity of the detector at this spectral region. In any case, this new scenario marks the turning point of the phase change to poly-crystalline Si which is irreversible because once it appears, no peaks, even the one associated with the Si hydride emission, could be observed with decreasing the excitation intensity.

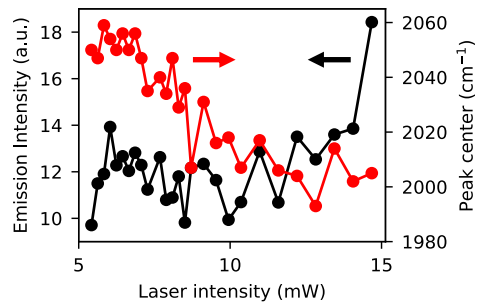

Fig 5: Integral (black dots) and center position (red dots) of the emission peak at 2000 cm$^{-1}$ at different laser excitation intensities, corresponding to the spectra of fig 4(a).

*3.3 Coupling of the phonon emission to resonant modes defined by the spherical microcavity.*

We learned from a previous study on poly-crystalline silicon microspheres that Mie resonant peaks originate mainly from free carrier (*FC*) emission. This process has an exponential dependence with temperature because it includes the contribution of the density of free carriers and their cross section [11,17]. We observed that temperatures higher than 500 ºC were required for being able to observe the resonant emission. As we pointed out, this does not mean that such *FC* emission did not occur at lower temperatures but it was not detectable, i.e. the signal was below the noise of the detection system. The scenario is different with a-Si:H microspheres because, according to the experiments, the phonon emission intensity due to Si hydrides is much higher than *FC* emission at least at temperatures below the phase transition, i. e. at 600 ºC. However, it is around these temperatures where the *FC* emission becomes comparable to the phonon emission. Therefore, it is not an easy task to find a point where both phenomena coexist and can be measured. In fact, we think that the point (black dot) in fig. 5 corresponding to the highest excitation intensity with notably higher emission intensity than the other ones includes the contribution of both effects: *FC* and Si hydrides emission because the phonon peak at 2000 cm$^{-1}$ is probably very close to a Mie mode and can couple to it.

In order to better asses the coupling of the phonon peak to Mie resonances, we recorded the thermal emission at a higher resolution than in the previous section and compare the spectra of microspheres in the amorphous state with those obtained after the poly-crystalline state had been achieved.

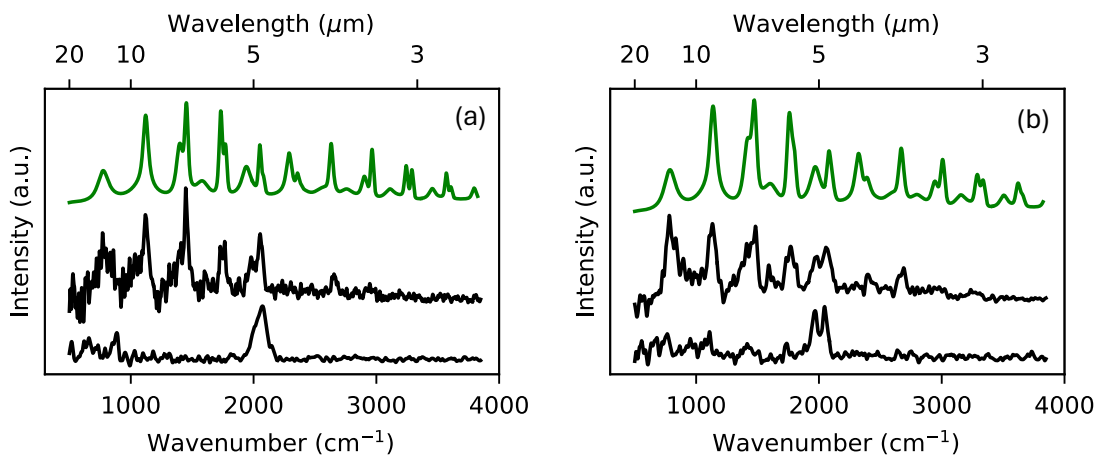

Fig 6: (a) Measured thermal emission spectra of a silicon microsphere before (bottom black curve) and after (middle black curve) the phase transition from amorphous to polycrystalline state. The green curve corresponds to the simulated spectrum of a crystalline silicon microsphere of 3470 nm in diameter at 600 ºC. (b) The same as (a) but for another microsphere with diameter of 3420 nm. In both simulations appropriate convolution has been realized in order to account for the finite resolution of the measurements.

Figure 6 (a) and (b) show these measurements for two different microspheres (black curves), as well as Mie theory based thermal emission calculations for crystalline silicon spheres of 3470 and 3420 nm in diameter respectively at 600 ºC (green curves). In both cases, the phonon emission peak coincides with Mie resonances, perhaps more in the second case than in the first one where there is a partial overlapping. The emergence of a doublet with pronounced peaks in the second case [fig 6(b)] confirms that phonon emission couples in fact to two Mie modes. Here, we are assuming that small changes in the sphere diameter and in the refractive index coming from the phase transition should not produce big changes in the resonances position. This is a reasonable approach because we are studying a region with low number Mie modes, i.e. modes with low $Q$ (quality factor). Anyhow, it is not trivial that the crystallization by laser irradiation preserves the spherical shape of the particles. Although we studied the structural changes occurred during this transition in annealing processes by electron microscopy techniques [18], the emergence of Mie resonances constitutes a very reliable sign of their good structural quality.

## 4. Conclusion

The thermal emission of light in the mid infrared of a-Si:H microspheres comes from vibrations of Si hydrides bonds, principally Si-H. The emission consists mainly of a peak centred around 2000 cm$^{-1}$ and it could couple to Mie resonances favoured by the spherical cavity for appropriate sphere diameters.

The first exposure of a microsphere to the light of a laser in the range of high absorption, namely at $\lambda$= 405 nm, produces notable changes in the scattering spectrum at VIS-nearIR wavelengths that we have attributed to variations of sphere diameter and refractive index. After that, a more stable stage is achieved and a quite stable emission could be recorded during more than one hour, although there seems to occur a continuous rearrangement of H atoms, probably from hydrides at the surface of micro or nano-voids existing in the Si amorphous matrix.

a-Si:H microspheres could be successfully converted into poly-crystalline ones by increasing the mentioned excitation laser to a certain threshold. This change is not reversible and it is characterized by the disappearance of the Si hydrides emission and the emergence of several peaks corresponding to Mie resonances that are originated from free carrier emission.

**Acknowledgements**

This research was supported by the project MCIN/AEI/PID2021-123163OB-I00, and